\documentclass[english,nolineno,a4paper]{socg-lipics-v2021}
\hideLIPIcs
\usepackage{algorithm}
\usepackage{amsmath}
\usepackage{mathtools}
\usepackage{cite}
\usepackage{algpseudocode}
\newcommand{\etal}{\textit{et al.}}

\DeclareMathOperator{\E}{E}
\DeclareMathOperator{\D}{\mathcal{D}}
\DeclareMathOperator{\Or}{\mathcal{O}}
\usepackage{xcolor}
\newcommand{\adi}[1]{}
\newcommand{\dave}[1]{}

\title{Tracking Evolving Labels using Cone based Oracles\footnote[0]{This is an abstract of a presentation given at CG:YRF 2023. It has been made public for the benefit of the community and should be considered a preprint rather than a formally reviewed paper. Thus, this work is expected to appear in a conference with formal proceedings and/or in a journal.}} %TODO Please add

\titlerunning{Tracking Evolving Labels using Cone based Oracles} %TODO optional, please use if title is longer than one line

\author{Aditya Acharya}{Department of Computer Science\\University of Maryland, College Park MD, USA}{adach@umd.edu}{https://orcid.org/0000-0002-0359-1913}{}

\author{David M. Mount}{Department of Computer Science and Institute for Advanced Computer Studies\\University of Maryland, College Park MD, USA}{mount@umd.edu}{https://orcid.org/0000-0002-3290-8932}{}%TODO mandatory, please use full name; only 1 author per \author macro; first two parameters are mandatory, other parameters can be empty. Please provide at least the name of the affiliation and the country. The full address is optional. Use additional curly braces to indicate the correct name splitting when the last name consists of multiple name parts.
\authorrunning{A. Acharya and D. M. Mount}
\Copyright{A. Acharya and D. M. Mount}

 %TODO mandatory: Please choose ACM 2012 classifications from https://dl.acm.org/ccs/ccs_flat.cfm 

 %TODO mandatory; please add comma-separated list of keywords

\category{} %optional, e.g. invited paper

\relatedversion{} %optional, e.g. full version hosted on arXiv, HAL, or other respository/website
\EventEditors{John Q. Open and Joan R. Access}
\EventNoEds{2}
\EventLongTitle{}
\EventShortTitle{}
\EventAcronym{}
\EventYear{}
\EventDate{}
\EventLocation{}
\EventLogo{}
\SeriesVolume{}
\ArticleNo{}
\begin{document}

\maketitle

%-----------------------------------------------------------------------
% \begin{abstract}
% %
% Motivated by the problem of maintaining data structures for a large sets of points that are evolving over the course of time, we consider the problem of maintaining a set of labels assigned to nodes in the Euclidean plane, where the locations of these labels change over time. We present a randomized algorithm based on a simple routing scheme, using cone based oracles, which maintains labels to within an average distance of $O(1)$ of their actual locations in expectation.
% \end{abstract}
% %-----------------------------------------------------------------------
\section{Introduction}
\label{sec:intro}

The \emph{evolving data framework} was first proposed by Anagnostopoulos {\etal}~\cite{evolving_sorting_first}, where an \emph{evolver} makes small changes to a structure behind the scenes. Instead of taking a single input and producing a single output, an algorithm judiciously \emph{probes} the current state of the structure and attempts to continuously maintain a sketch of the structure that is as close as possible to its actual state. There have been a number of problems that have been studied in the evolving framework \cite{evolving_sorting, evolving_graph, evolving_stable_matching, page_rank} including our own work on labeled trees \cite{CCCGtree}.  
%Acharya, and Mount \cite{CCCGtree}, later modified this framework to make it suitable for geometric data sets, which we restrict ourselves to in this paper.

We were motivated by the problem of maintaining a labeling in the plane, where updating the labels require physically moving them. Applications involve tracking evolving disease hot-spots via mobile testing units \cite{hotspot}, and tracking unmanned aerial vehicles \cite{uav}.

To be specific, we consider the problem of tracking labeled nodes in the plane, where an evolver continuously swaps labels of any two nearby nodes in the background unknown to us. We are tasked with maintaining a hypothesis, an approximate sketch of the locations of these labels, which we can only update by physically moving them over a sparse graph.  We assume the existence of an \emph{Oracle}, which when suitably probed, guides us in fixing our hypothesis.

\section{Theta Graphs and Routing}
\begin{figure}[h]
\centering
\includegraphics[width = 0.9\linewidth]{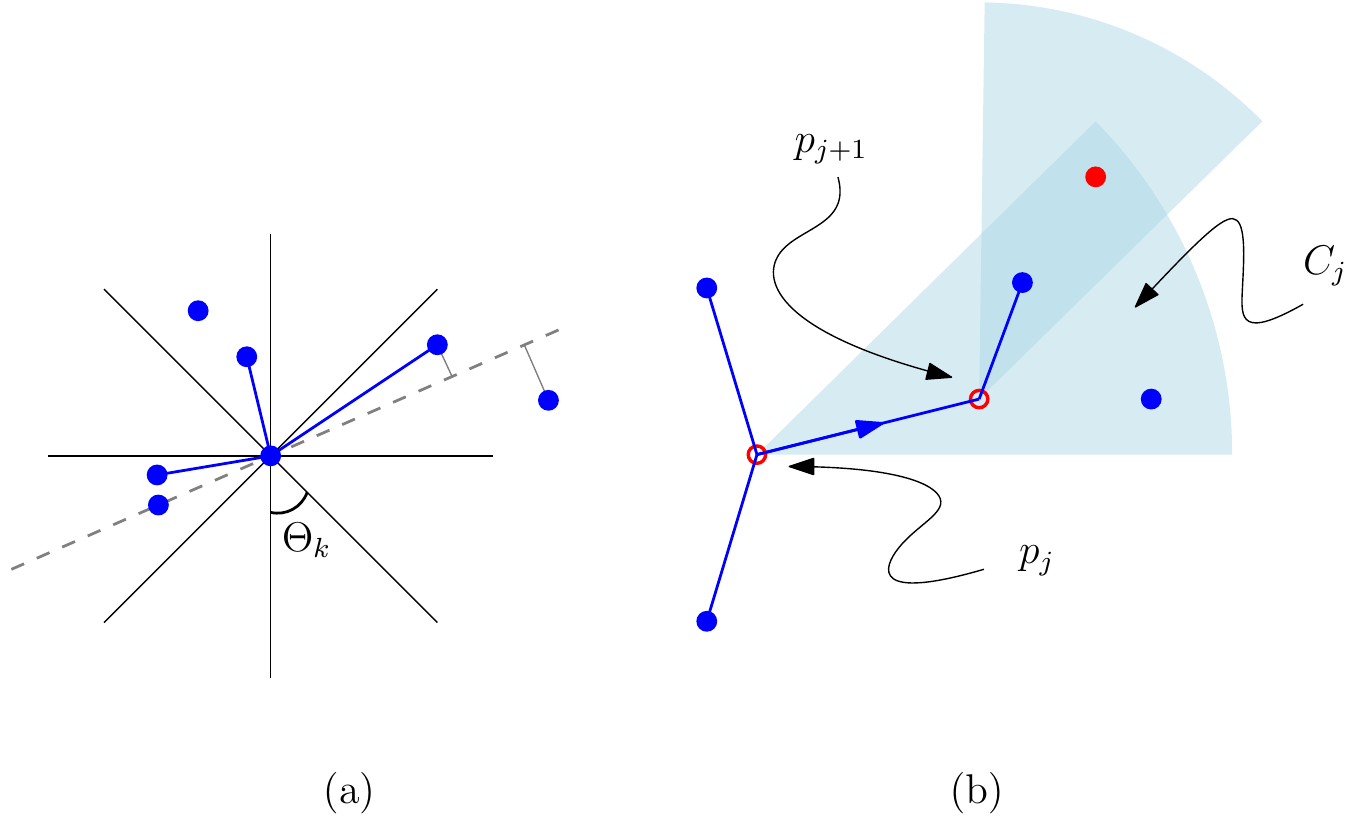}

\caption{Theta Graph: Construction, and Routing. 1(a) shows the construction of theta graphs. For a point, consider the set of cones around it. For every cone add an edge between that point, and the point which is nearest to it when projected along the bisector of the cone. 1(b) shows how to route using theta graphs. If $C_j$ is the cone at $p_j$ containing the destination point (Red pt), then follow the edge contained inside it. Repeat the same at $p_{j+1}$ until destination is reached}
\label{fig:theta}
\end{figure}

Theta graph is a sparse geometric spanner with a spanning ratio:  $t_\theta = 1/(1-2\sin(\theta_k /2))$ spanners, where $\theta_k = 2\pi/k$ is the angle induced by a set of cones around a point \cite{theta_span} (Figure \ref{fig:theta}(a)). Using a pre-computed theta graph, there is a simple routing scheme: let $C_{p,q}$ be the cone around $p$ that contains $q$. We move from $p$ to its neighbor in $C_{p,q}$, and repeat the process. (See Figure \ref{fig:theta}(b)). Note we only need information about the neighbors of $p$, and a rough location estimate for the target point $q$.

\section{Problem Formulation}
Let $P \in \mathbf{R}^2$, be a point set of size $n$. Each point is assigned a unique label from the set of labels $L = \{l_1, \ldots, l_n\}$, through a matching $M$. Every time step, the evolver swaps labels between points that are less than a unit distance apart.

Let $G_\theta$ be the theta graph on $P$, and $\Sigma_\theta$ be its embedding. Through $H: L \rightarrow \Sigma_\theta$, we maintain a mapping as close as possible to the original one. We update $H$ by moving a single label over $\Sigma_{\theta}$ with a finite speed, at any particular moment of time. One of our goals is to find such a speed, which is sufficient to maintain a reasonable $H$.

We define the distance $\mathcal D _l$, for label $l$ as the Euclidean distance between its hypothesized location, and its actual location. And the overall distance between the two mappings $M$, and $H$ as the sum of all such distances: $\sum _{l\in L} \mathcal D _l = \mathcal D(M,H)$. It's easy to see that, this can get as large as $O(n^2)$ without a competing algorithm.

We assume we have access to an oracle, that we call a \emph{Cone Oracle}: $\Or$, which when queried on a label $l$, and its hypothesized location $H(l)$, returns the cone $C_k$ around $H(l)$ which contains the true location of $l$. 
%Physically, one can think of the oracle as an antenna having a directional spread that provides us the rough sub-space that contains the actual label. 

We assume an additional thing here: for a label $l$, after an initial startup cost, subsequent queries on the oracle with the same label, takes $o(1)$ time. This is a valid assumption on massive point sets, when only a small region around the current point, in our case $H(l)$, is \emph{cached} in memory. The expensive step is moving to a completely different, possibly uncached area of the data set. In other words switching from $l_1$ to $l_2$, on the oracle takes constant amount of time, what we call the \emph{memory overhead}. 
%These kind of caching techniques, where the graphs are stored in tiers of memory is frequently seen in practice \cite{cachedpath}\cite{bulucc2017distributed}.

Our aim is to design an algorithm that physically moves the hypothesized labels over a sparse graph with a constant speed, and maintains a labeling of distance $O(n)$ at all times.
%In the next section we show there exists a randomized algorithm that moves labels at the speed of $3\,t_\theta$ over a Theta graph, with $k\geq 7$, and maintains a labeling of distance $O(n)$, in expectation.

\section{Algorithm}
First we precompute the theta graph for our point set $P$. Then we run the randomized algorithm \ref{alg:theta}. (We assume the evolver does not have access to our random bit generator). In short, our algorithm selects a label at random every iteration, and tracks it down using the routing scheme on theta graphs (Fig \ref{fig:theta}(b)). We probe the oracle to find the cone around a point which contains the actual location of the label.
%Rupert and Seidel \cite{theta_span} give an $O(n\log n)$ time algorithm to construct the theta graph in plane.The algorithm is described in Algorithm~\ref{alg:theta}. 
%Then, every iteration we choose a random label $l$, and track it's true location. We assume we have an oracle which returns the cone around $H(l)$ that contains the true location of $l$. Using the returned cone, we find the corresponding edge using our constructed theta graph, and move $l$ at a speed of $c \kern+1pt t_\theta$. We move on to another label, once the oracle returns NULL. We assume the evolver does not have access to our random bit generator.
\begin{algorithm}[h]
\caption{Randomized algorithm using cone oracle}\label{alg:theta}
\begin{algorithmic}[1]
\Require $\{H(l_1), H(l_2), \cdots, H(l_n)\}$ \Comment{Initial hypothesized location of the labels}
\While{true}\Comment{Continuously run the algorithm}
\For{$l$ chosen randomly uniformly from $\{l_1,l_2,\cdots,l_n\}$} 
\While{$ \mathcal{O}(l, H(l)) \neq NULL$} \Comment{Keep tracking $l$ until found}
    \State{Let $C_k \leftarrow \mathcal{O}(l, H(l))$}\Comment{Cone returned by the oracle}
    \State{Let $e_{l,k}= \left(H(l),v\right)$ be the edge incident on $H(l)$ inside $C_k$}
    \State{Move $l$ along $e_{l,k}$ with speed $c \kern+1pt t_\theta$} \Comment{$c$ is a constant}
    \State{Update $H(l) \leftarrow v$}
\EndWhile
\EndFor
\EndWhile
\end{algorithmic}
\end{algorithm}

\section{Analysis}
%Let the memory overhead be $M$, 
%or in other words it takes $M$ units of time to change a label and begin tracking it over the graph. 
Under a cached memory model, with overhead $M$,
%where the oracle can operate in parallel, 
lines 3, 4, 5, and 7 in Algorithm~\ref{alg:theta} take $o(1)$ time. 
%We define  a single iteration as picking a random label and tracking it, that is, a single loop of the \emph{for loop} in line 2. 
Let $\mathcal{D}_i$ be the overall distance, and $\mathcal{D}_{l,i}$ be the distance for label $l$ at the start of the $i^{th}$ iteration. Since we are moving at the speed of $c \kern+1pt t_\theta$ over a $t_\theta$ spanner, we spend $\mathcal{D}_{l,i} /c$ time in traversing Euclidean distance $\mathcal{D}_{l,i}$. In that time the evolver could have moved the label by at most another $\mathcal{D}_{l,i}/{c}$ distance. So, in the worst case, we spend at most $\mathcal{D}_{l,i} /c + \mathcal{D}_{l,i} /{c^2} + \cdots = \mathcal{D}_{l,i}/(c-1)$ time tracking $l$. In that time, the evolver can increase $\mathcal{D}_i$ by at most $2\mathcal{D}_{l,i}/(c-1)$. 
%The algorithm reduced $\mathcal{D}_i$ by $D_l$ by completely tracking $l$. 
Since we chose $l$ at random, we have $\E [\mathcal{D}_{l,i}] = \mathcal{D}_i /n$. Therefore,
\begin{align*}
    \E\left[\D_{i+1} \mid \D_i\right] & ~ \leq ~ \D_i - \frac{\D_i}{n} + \frac{2}{c-1}\cdot \frac{\D_i}{n} + M &\\
    \implies \E\left[\D_{i+1}\right] & ~ \leq ~ \left(1 - \frac{c-3}{c-1}\cdot \frac{1}{n}\right)\E\left[\D_i\right] + M &\\
    %& \leq \left(1-\frac{z}{n}\right)\E\left[\D_i\right] + M &\text{$\left[z = \frac{c-3}{c-1}\right]$}\\
    & \leq \left(1-\frac{z}{n}\right)^i\E\left[\D_1\right] + M\sum_{j=1}^{i}\left(1-\frac{z}{n}\right)^i&\text{$\left[z = \frac{c-3}{c-1}\right]$}\\
    & \leq \left(1-\frac{z}{n}\right)^i O(n^2) + \frac{nM}{z} &\text{$\left[z > 0\right]$}\\
\end{align*}

For $i = \ln n /(\ln n - \ln (n-z)) \sim n\ln n/z $, we have $\D_{i+1} \in O(n)$, provided 
%$z = (c-3)/(c-1) > 0$, which implies 
$c >3$. Therefore, 
%Therefore a movement speed of $3\,t_\theta$ is sufficient for our purposes.

\begin{theorem}
There exists a randomized algorithm using a cone oracle, which moves labels at any speed greater than $3\,t_\theta$, and in expectation maintains a hypothesized labeling with distance: $O(n)$, in the presence of an adversarial evolver.
%which does not have access to the random bits. 
\end{theorem}

%-----------------------------------------------------------------------
\bibliography{evolving}
%-----------------------------------------------------------------------

\end{document}